\documentstyle[12pt]{article}

\begin{document}

\title{\vskip-3cm{\baselineskip14pt
\centerline{\normalsize DESY 99-121\hfill ISSN 0418-9833}
\centerline{\normalsize KEK-TH-638\hfill}
\centerline{\normalsize hep-ph/9909502\hfill}
\centerline{\normalsize September 1999\hfill}
}
\vskip1.5cm
Quark-Loop Amplitudes for $W^\pm H^\mp$ Associated Hadroproduction
}
\author{A.A. Barrientos Bendez\'u$^1$ and B.A. Kniehl$^{2,}$\thanks{Permanent
address: II. Institut f\"ur Theoretische Physik, Universit\"at Hamburg,
Luruper Chaussee 149, 22761 Hamburg, Germany.}\\
{\normalsize $^1$
II. Institut f\"ur Theoretische Physik, Universit\"at Hamburg,}\\
{\normalsize Luruper Chaussee 149, 22761 Hamburg, Germany}\\
{\normalsize $^2$
High Energy Accelerator Research Organization (KEK), Theory Division,}\\
{\normalsize 1-1 Oho, Tsukuba-shi, Ibaraki-ken, 305-0801 Japan}}

\date{}

\maketitle

\thispagestyle{empty}

\begin{abstract}
In this addendum to our paper entitled {\it $W^\pm H^\mp$ Associated Production
at the Large Hadron Collider} [Phys.\ Rev.\ D {\bf59}, 015009 (1999)], we list
analytic results for the helicity amplitudes of the partonic subprocess
$gg\to W^-H^+$ induced by virtual quarks.

\medskip

\end{abstract}

\newpage

In a recent paper \cite{bar}, we studied the hadroproduction of a charged Higgs
boson in association with a $W$ boson at the CERN Large Hadron Collider (LHC) in
the context of the two-Higgs-doublet model of type~II, which serves as the Higgs
sector for the minimal supersymmetric extension of the standard model (SM).
This reaction dominantly proceeds via the partonic subprocesses
$b\bar b\to W^\pm H^\mp$ at the tree level (see Fig.~1 in Ref.~\cite{bar}) and
$gg\to W^\pm H^\mp$, which is mediated by triangle- and box-type diagrams
involving virtual top and bottom quarks (see Fig.~2 in Ref.~\cite{bar}).
In Ref.~\cite{bar}, we presented analytic expressions for the cross section of
$b\bar b\to W^\pm H^\mp$ and the transition-matrix element of
$gg\to W^\pm H^\mp$ arising from the quark triangles.
However, we refrained from listing our formulas for the quark box contributions
because we found that were somewhat lengthy.
In the meantime, a signal-versus-background analysis of $W^\pm H^\mp$
associated production at the LHC was carried out by Moretti and Odagiri
\cite{mor}, who generated the signal cross section by using the formulas
published in Ref.~\cite{bar}, thus omitting the quark box contributions.
This motivated us to further compactify our expressions for the latter by
introducing helicity amplitudes.
The purpose of this brief report is to provide these results, which may be
useful for other authors as well.

Calling the four-momenta of the two gluons and the $W$ boson $p_a$, $p_b$, and
$p_W$, respectively, we define the partonic Mandelstam variables as
$s=(p_a+p_b)^2$, $t=(p_a-p_W)^2$, and $u=(p_b-p_W)^2$.
Furthermore, we introduce the following short-hand notations: $w=m_W^2$,
$h=m_H^2$, $d=t-u$, $t_1=t-w$, $t_2=t-h$, $u_1=u-w$, $u_2=u-h$, $N=tu-wh$,
$\lambda=s^2+w^2+h^2-2(sw+wh+hs)$, and $q=m_t^2-m_b^2$.
We label the helicity states of the two gluons and the $W$ boson in the partonic
center-of-mass frame by $\lambda_a=-1/2,1/2$, $\lambda_b=-1/2,1/2$, and
$\lambda_W=-1,0,1$.
For reference, we first list the helicity amplitudes for the quark triangle
contributions, ${\cal M}^\triangle_{\lambda_a\lambda_b\lambda_W}$.
They may be extracted from Eq.~(5) of Ref.~\cite{bar} and read
\begin{equation}
{\cal M}^\triangle_{\lambda_a\lambda_b0}=\frac{s\sqrt\lambda}{m_W}
[(1+\lambda_a\lambda_b)\Sigma(s)-(\lambda_a+\lambda_b)\Pi(s)],
\end{equation}
where $\Sigma$ and $\Pi$ are the vector and axial-vector form factors given in
Eq.~(6) of Ref.~\cite{bar}.
In this case, the $W$ boson can only be longitudinally polarized because it
couples to two Higgs bosons, so that
${\cal M}^\triangle_{\lambda_a\lambda_b\lambda_W}=0$ for $\lambda_W=\pm1$.
As for the quark box contributions, all twelve helicity amplitudes,
${\cal M}_{\lambda_a\lambda_b\lambda_W}^\Box$, contribute.
Due to Bose and weak-isospin symmetry, they are related by
\begin{eqnarray}
{\cal M}_{\lambda_a\lambda_b0}^\Box\left(t,u,m_b^2,m_t^2,\tan\beta\right)&=&
{\cal M}_{\lambda_b\lambda_a0}^\Box\left(u,t,m_b^2,m_t^2,\tan\beta\right),
\nonumber\\
{\cal M}_{\lambda_a\lambda_b\lambda_W}^\Box\left(t,u,m_b^2,m_t^2,\tan\beta
\right)&=&
-{\cal M}_{\lambda_b\lambda_a\lambda_W}^\Box\left(u,t,m_b^2,m_t^2,\tan\beta
\right),
\nonumber\\
{\cal M}_{\lambda_a\lambda_b0}^\Box\left(t,u,m_b^2,m_t^2,\tan\beta\right)&=&
-{\cal M}_{-\lambda_a-\lambda_b0}^\Box\left(t,u,m_t^2,m_b^2,\cot\beta\right),
\nonumber\\
{\cal M}_{\lambda_a\lambda_b\lambda_W}^\Box\left(t,u,m_b^2,m_t^2,\tan\beta
\right)&=&
{\cal M}_{-\lambda_b-\lambda_a-\lambda_W}^\Box\left(u,t,m_t^2,m_b^2,\cot\beta
\right).
\label{eq:m}
\end{eqnarray}
Keeping $\lambda_W=\pm1$ generic, we thus only need to specify four expressions.
These read:
\begin{eqnarray}
{\cal M}_{++0}^\Box&=&\frac{2}{m_Ws\sqrt\lambda}
\left[\left(m_b^2\tan\beta+m_t^2\cot\beta\right)F_{++}^0+m_t^2\cot\beta G_{++}^0
+(t\leftrightarrow u)\right],
\nonumber\\
{\cal M}_{+-0}^\Box&=&\frac{1}{m_WN\sqrt\lambda}
\left[\left(m_b^2\tan\beta+m_t^2\cot\beta\right)F_{+-}^0+m_t^2\cot\beta G_{+-}^0
\right.
\nonumber\\
&&{}-\left.\left(t\leftrightarrow u,m_b^2\leftrightarrow m_t^2,
\tan\beta\leftrightarrow\cot\beta\right)\right],
\nonumber\\
{\cal M}_{++\lambda_W}^\Box&=&\sqrt{\frac{2}{sN}}
\left[\frac{m_b^2\tan\beta+m_t^2\cot\beta}{s}
\left(\frac{F_{++}^1}{\sqrt\lambda}+\lambda_WF_{++}^2\right)
+m_t^2\cot\beta\left(\frac{G_{++}^1}{\sqrt\lambda}+\lambda_WG_{++}^2\right)
\right.
\nonumber\\
&&{}-\left.\vphantom{\frac{F_{++}^1}{\sqrt\lambda}}(t\leftrightarrow u)\right],
\nonumber\\
{\cal M}_{+-\lambda_W}^\Box&=&\frac{1}{\sqrt{2sN}}
\left[\frac{m_b^2\tan\beta+m_t^2\cot\beta}{N}
\left(\frac{F_{+-}^1}{\sqrt\lambda}+\lambda_WF_{+-}^2\right)
+m_t^2\cot\beta\left(\frac{G_{+-}^1}{\sqrt\lambda}+\lambda_WG_{+-}^2\right)
\right.
\nonumber\\
&&{}+\left.\vphantom{\frac{F_{+-}^1}{\sqrt\lambda}}
\left(t\leftrightarrow u,m_b^2\leftrightarrow m_t^2,
\tan\beta\leftrightarrow\cot\beta,\lambda_W\rightarrow-\lambda_W\right)\right], 
\end{eqnarray}
where $F_{+\pm}^i$ and $G_{+\pm}^i$, with $i=0,1,2$, are complex functions of
$t$, $u$, $m_b^2$, and $m_t^2$.
The normalization of ${\cal M}_{\lambda_a\lambda_b\lambda_W}^\triangle$ and
${\cal M}_{\lambda_a\lambda_b\lambda_W}^\Box$ is such that the differential
cross section of $gg\to W^-H^+$ is given by
\begin{equation}
\frac{d\sigma}{dt}(gg\to W^-H^+)=\frac{\alpha_s^2(\mu)G_F^2m_W^2}
{256(4\pi)^3s^2}\sum_{\lambda_a,\lambda_b,\lambda_W}\left|
{\cal M}_{\lambda_a\lambda_b\lambda_W}^\triangle+
{\cal M}_{\lambda_a\lambda_b\lambda_W}^\Box\right|^2.
\label{eq:x}
\end{equation}

We now express the form factors $F_{+\pm}^i$ and $G_{+\pm}^i$ in terms of the 
standard scalar two-, three-, and four-point functions,
\begin{eqnarray}
\lefteqn{B_0\left(p_1^2,m_0^2,m_1^2\right)
=\int\frac{d^Dq}{i\pi^2}\,\frac{1}{\left(q^2-m_0^2+i\epsilon\right)
\left[(q+p_1)^2-m_1^2+i\epsilon\right]},}
\nonumber\\
\lefteqn{C_0\left(p_1^2,(p_2-p_1)^2,p_2^2,m_0^2,m_1^2,m_2^2\right)}
\nonumber\\
&=&\int\frac{d^Dq}{i\pi^2}\,\frac{1}{\left(q^2-m_0^2+i\epsilon\right)
\left[(q+p_1)^2-m_1^2+i\epsilon\right]\left[(q+p_2)^2-m_2^2+i\epsilon\right]},
\nonumber\\
\lefteqn{D_0\left(p_1^2,(p_2-p_1)^2,(p_3-p_2)^2,p_3^2,p_2^2,(p_3-p_1)^2,
m_0^2,m_1^2,m_2^2,m_3^2\right)}
\nonumber\\
&=&\int\frac{d^Dq}{i\pi^2}\,\frac{1}{\left(q^2-m_0^2+i\epsilon\right)
\left[(q+p_1)^2-m_1^2+i\epsilon\right]\left[(q+p_2)^2-m_2^2+i\epsilon\right]
\left[(q+p_3)^2-m_3^2+i\epsilon\right]},
\nonumber\\
\end{eqnarray}
where $D$ is the space-time dimensionality.
The $B_0$ function is ultraviolet (UV) divergent in the physical limit $D\to4$,
while the $C_0$ and $D_0$ functions are UV finite in this limit.
We evaluate the $B_0$, $C_0$, and $D_0$ functions numerically with the aid of
the program package FF \cite{old}.
To simplify to notation, we introduce the abbreviations
$C_{ijk}^{ab}(c)=C_0\left(a,b,c,m_i^2,m_j^2,m_k^2\right)$ and
$D_{ijkl}^{abcd}(e,f)=D_0\left(a,b,c,d,e,f,m_i^2,m_j^2,m_k^2,m_l^2\right)$.
We find
\begin{eqnarray}
F_{++}^0&=&-2s(t_1+u_1)\left[m_b^2C^{00}_{bbb}(s)-m_t^2C^{00}_{ttt}(s)\right]
+f_1(t,u,q)\left[t_2C^{h0}_{btt}(t)+t_1C^{w0}_{tbb}(t)\right]
\nonumber\\
&&{}+f_1(u,t,q)\left[t_2C^{h0}_{tbb}(t)+t_1C^{w0}_{btt}(t)\right]
-f_1(t,u,q)\left[N+s(m_b^2+m_t^2)\right]D^{h0w0}_{bttb}(t,u)
\nonumber\\
&&{}-2sm_b^2[2wt_2+f_1(t,u,q)]D^{hw00}_{btbb}(s,t)
+2sm_t^2[2wu_2-f_1(t,u,q)]D^{hw00}_{tbtt}(s,t),
\nonumber\\
G_{++}^0&=&-2s^2(t+u)C^{00}_{ttt}(s)
-t_2f_1(-t_2,u_2,h)\left[C^{h0}_{btt}(t)+C^{h0}_{tbb}(t)\right]
-t_1f_1(-t_2,u_2,h)
\nonumber\\
&&{}\times\left[C^{w0}_{btt}(t)+C^{w0}_{tbb}(t)\right]
+\left[2s\lambda m_b^2+(N+sq)f_1(-t_2,u_2,h)\right]D^{h0w0}_{bttb}(t,u)
\nonumber\\
&&{}+2s\lambda m_b^2D^{hw00}_{btbb}(s,t)
+2s\left[2swh-sm_b^2(t+u)+m_t^2f_1(-t_2,u_2,h)\right]D^{hw00}_{tbtt}(s,t),
\nonumber\\
F_{+-}^0&=&2s\{2wN+(t+u)[N-f_2(0,t,\lambda)]+qf_2(t,u,2\lambda)+2q^2(t_1+u_1)\}  
C^{00}_{bbb}(s)
\nonumber\\
&&{}-2t_2[wh(t_1+u_1)+qf_1(2u,2h,t)]C^{h0}_{btt}(t)
+2t_2[w(hd-2tt_2-2N)
\nonumber\\
&&{}-qf_1(2u,2h,t)]C^{h0}_{tbb}(t)
-2t_1[w(hd-2ut_2)+qf_1(2t,2h,t)]C^{w0}_{btt}(t)
\nonumber\\
&&{}+2t_1[w(hd-2tt_2)-qf_1(2t,2h,t)]C^{w0}_{tbb}(t)
-2\{\lambda(ud-2wt_2)-(t_1+u_1)
\nonumber\\
&&{}\times[N(t+u)+q(d^2+2N)]\}C^{hw}_{btb}(s)
-[ud-2N-q(t_1+u_1)]\!\!\left[N\!\left(m_b^2+m_t^2\right)+sq^2\right]
\nonumber\\
&&{}\times D^{h0w0}_{bttb}(t,u)
-\left\{2N\left[wN-m_b^2f_3(t,u,2w)\right]
-Nq\left[ud+2u_1^2+2m_b^2(t_1+u_1)\right]\right.
\nonumber\\
&&{}-\left.q^2[du_1(t_1+u_1)-t_2f_3(t,u,w)]-sq^3(t_1+u_1)\right\}
D^{h0w0}_{tbbt}(t,u)
-2\left\{stw(hd-2tt_2)\right.
\nonumber\\
&&{}-2Nm_b^2f_2(0,t,\lambda)
-q\left[sN(t+u)-stf_2(t,0,2\lambda)+2Nm_b^2(t_1+u_1)\right]
\nonumber\\
&&{}-\left.sq^2f_2(t,0,\lambda)-sq^3(t_1+u_1)\right\}D^{hw00}_{btbb}(s,t)
+2(t_1+u_1)\left\{stwh+2uNm_t^2\right.
\nonumber\\ 
&&{}+\left.q\left[st(t+2u)-N\left(s-2m_t^2\right)\right]
+sq^2(2t+u)+sq^3\right\}
D^{hw00}_{tbtt}(s,t),
\nonumber\\
G_{+-}^0&=&2s\left[(t+u)(d^2+2N)-2\lambda q\right]C^{00}_{bbb}(s)  
+2\left[t^2(t+u)-wh(3t-u)\right]
\nonumber\\
&&{}\times\left\{t_2\left[C^{h0}_{btt}(t)+C^{h0}_{tbb}(t)\right] 
+t_1\left[C^{w0}_{btt}(t)+C^{w0}_{tbb}(t)\right]\right\} 
-2\lambda(d^2+2N)C^{hw}_{btb}(s)
\nonumber\\
&&{}-\lambda\left[N\left(m_b^2+m_t^2\right)+sq^2\right]D^{h0w0}_{bttb}(t,u)
+\left[2N\left(2wN-\lambda m_b^2\right)-\lambda q(N+sq)\right]
\nonumber\\
&&{}\times D^{h0w0}_{tbbt}(t,u)
-2f_4\left(m_b^2,m_t^2\right)D^{hw00}_{btbb}(s,t)
-2f_4\left(m_t^2,m_b^2\right)D^{hw00}_{tbtt}(s,t),
\nonumber\\
F_{++}^1&=&2s^2d\left[m_b^2C^{00}_{bbb}(s)-m_t^2C^{00}_{ttt}(s)\right]
-f_5\left(w,m_b^2,m_t^2\right)\left[t_2C^{h0}_{btt}(t)+t_1C^{w0}_{tbb}(t)\right]
\nonumber\\
&&{}+f_5\left(w,m_t^2,m_b^2\right)
\left[t_2C^{h0}_{tbb}(t)+t_1C^{w0}_{btt}(t)\right]
+\left[N+s(m_b^2+m_t^2)\right]f_5\left(w,m_b^2,m_t^2\right)
\nonumber\\
&&{}\times D^{h0w0}_{bttb}(t,u)
+2sf_6\left(m_b^2,m_t^2\right)D^{hw00}_{btbb}(s,t)
-2sf_6\left(m_t^2,m_b^2\right)D^{hw00}_{tbtt}(s,t),
\nonumber\\
F_{++}^2&=&-(N-sq)\left[t_2C^{h0}_{btt}(t)-t_1C^{w0}_{btt}(t)\right]
+(N+sq)\left[t_2C^{h0}_{tbb}(t)-t_1C^{w0}_{tbb}(t)\right]
\nonumber\\
&&{}+\left\{N\left[N+2s(m_b^2+m_t^2)\right]+s^2q^2\right\}D^{h0w0}_{bttb}(t,u),
\nonumber\\
G_{++}^1&=&-sd(t_2+u_2)C^{00}_{ttt}(s)
+t_2f_3(t,u,h)\left[C^{h0}_{btt}(t)+C^{h0}_{tbb}(t)\right]
-t_1f_3(u,t,h)
\nonumber\\
&&{}\times\left[C^{w0}_{btt}(t)+C^{w0}_{tbb}(t)\right]
+(N+sq)f_3(u,t,h)D^{h0w0}_{bttb}(t,u)
+s(t_2+u_2)[2N+d(t+q)]
\nonumber\\
&&{}\times D^{hw00}_{tbtt}(s,t),
\nonumber\\
G_{++}^2&=&-sdC^{00}_{ttt}(s) 
+t_2u_2\left[C^{h0}_{btt}(t)+C^{h0}_{tbb}(t)\right]
+(st+N)C^{w0}_{btt}(t)-t_1t_2C^{w0}_{tbb}(t)
\nonumber\\
&&{}+t_2(N+sq)D^{h0w0}_{bttb}(t,u)+s[2N+d(t+q)]D^{hw00}_{tbtt}(s,t),
\nonumber\\
F_{+-}^1&=&-4sdNB_0\left(s,m_b^2,m_b^2\right)
-2s\{sd(t^2+u^2)-N[4su+d(w-h)+\lambda]
\nonumber\\
&&{}-2sq[d(t+u)-2N]+2sdq^2\}C^{00}_{bbb}(s)
+2t_2\left[2sN(t+u)+f_7\left(m_b^2,m_t^2\right)\right]C^{h0}_{btt}(t)
\nonumber\\
&&{}-2t_2f_7\left(m_t^2,m_b^2\right)C^{h0}_{tbb}(t)
-2t_1\left[2t_2N(t_1+u_1)+f_8\left(m_b^2,m_t^2\right)\right]C^{w0}_{btt}(t)
\nonumber\\
&&{}+2t_1f_8\left(m_t^2,m_b^2\right)C^{w0}_{tbb}(t)
-2s\lambda[N-d(t+u-2q)]C^{hw}_{btb}(s)
+f_5\left(h,m_t^2,m_b^2\right)
\nonumber\\
&&{}\times\left[N\left(m_b^2+m_t^2\right)+sq^2\right]
D^{h0w0}_{bttb}(t,u)
-\left\{N^2\left[f_3(t,u,w)+2m_t^2(3s+w-h)\right]\right.
\nonumber\\
&&{}-\left.Nq\left[s\lambda-N(t_1+u_1)-sd\left(t_1+2m_t^2\right)\right]
+2sNq^2(2s+u_2)+s^2dq^3\right\}D^{h0w0}_{tbbt}(t,u)
\nonumber\\
&&{}-2s(t-q)\left\{Nf_3(u,t,h)-d\left[st(t-2m_t^2)-2t_1t_2m_b^2+sq^2\right]
\right\}D^{hw00}_{btbb}(s,t) 
\nonumber\\
&&{}+2s\left\{stdN-(ud+\lambda)\left(st^2+2Nm_t^2\right)
-q\left[sd(t(t+2u)-N)+2st\lambda+2dNm_t^2\right]\right.
\nonumber\\
&&{}-\left.sq^2[d(2t+u)+\lambda]-sdq^3\right\}D^{hw00}_{tbtt}(s,t),
\nonumber\\
F_{+-}^2&=&-4sNB_0\left(s,m_b^2,m_b^2\right)
-2s\left[s(t^2+u^2)+N(t_2-u_1)+4Nm_b^2-2sq(t+u)+2sq^2\right]
\nonumber\\
&&{}\times C^{00}_{bbb}(s)
-2t_2\left[u_1N-f_9\left(t_2,m_b^2,m_t^2\right)\right]C^{h0}_{btt}(t)
-2t_2\left[u_1N+f_9\left(-t_2,m_t^2,m_b^2\right)\right]
\nonumber\\
&&{}\times C^{h0}_{tbb}(t)
+2t_1f_9\left(t_2,m_b^2,m_t^2\right)C^{w0}_{btt}(t)
-2t_1f_9\left(-t_2,m_t^2,m_b^2,\right)C^{w0}_{tbb}(t)
+2s[d^2(t+u)
\nonumber\\
&&{}+N(t+3u)-2q(d^2+2N)]C^{hw}_{btb}(s)
-q\left[N^2+2sN\left(m_b^2+m_t^2\right)+s^2q^2\right]D^{h0w0}_{bttb}(t,u)
\nonumber\\
&&{}-\left[u_1N^2+Nq\left(tu_1+ut_2+u_1^2+4sm_t^2\right)-2sNq^2+s^2q^3\right]
D^{h0w0}_{tbbt}(t,u)
\nonumber\\
&&{}-2f_{10}\left(t_2,m_b^2,m_t^2\right)D^{hw00}_{btbb}(s,t) 
+2f_{10}\left(s,m_t^2,m_b^2\right)D^{hw00}_{tbtt}(s,t),
\nonumber\\
G_{+-}^1&=&-2f_3(u,t,h)\left\{2sC^{00}_{bbb}(s)
+t_1\left[C^{w0}_{btt}(t)+C^{w0}_{tbb}(t)\right]\right\}
+2t_2f_3(t,u,h)\left[C^{h0}_{btt}(t)+C^{h0}_{tbb}(t)\right]
\nonumber\\
&&{}-[dN(t_1+u_1)+s\lambda q]D^{h0w0}_{tbbt}(t,u)
+2s(t-q)f_3(u,t,h)D^{hw00}_{btbb}(s,t)
\nonumber\\
&&{}-2s(t+q)f_3(t,u,h)D^{hw00}_{tbtt}(s,t),
\nonumber\\
G_{+-}^2&=&2t_2\left\{2sC^{00}_{bbb}(s)
+u_2\left[C^{h0}_{btt}(t)+C^{h0}_{tbb}(t)\right] 
+t_1\left[C^{w0}_{btt}(t)+C^{w0}_{tbb}(t)\right]\right\}
\nonumber\\
&&{}+f_5\left(w,m_b^2,m_t^2\right)D^{h0w0}_{tbbt}(t,u) 
-2st_2(t-q)D^{hw00}_{btbb}(s,t)-2su_2(t+q)D^{hw00}_{tbtt}(s,t),
\end{eqnarray}
where we have used the auxiliary functions
\begin{eqnarray}
f_1(t,u,q)&=&-w(t-u)+q(t_1+u_1),
\nonumber\\
f_2(t,u,\lambda)&=&\lambda-(3t+u)(t_1+u_1),
\nonumber\\
f_3(t,u,h)&=&2N-(t-u)(u-h),
\nonumber\\
f_4\left(m_b^2,m_t^2\right)&=&-st[N(t+u)-t\lambda]
-s(m_b^2-m_t^2)[N(t+u)-2t\lambda]
\nonumber\\
&&{}+\lambda\left[2Nm_b^2+s\left(m_b^2-m_t^2\right)^2\right],
\nonumber\\
f_5\left(w,m_b^2,m_t^2\right)&=&N(t+u-2w)-sd\left(m_b^2-m_t^2\right),
\nonumber\\
f_6\left(m_b^2,m_t^2\right)&=&-sm_b^2\left[2N+d\left(t+m_b^2-m_t^2\right)
\right],
\nonumber\\
f_7\left(m_b^2,m_t^2\right)&=&sd(t^2+N)
-\left(m_b^2-m_t^2\right)[2std+N(3s+w-h)],
\nonumber\\
f_8\left(m_b^2,m_t^2\right)&=&2N^2-d(st^2-t_2N)
+\left(m_b^2-m_t^2\right)[d(2st+N)-2t_2N],
\nonumber\\
f_9\left(t_2,m_b^2,m_t^2\right)&=&st^2-t_2N-\left(m_b^2-m_t^2\right)(2st+N),
\nonumber\\
f_{10}\left(t_2,m_b^2,m_t^2\right)&=&-s\left(t+m_b^2-m_t^2\right)
\left[st\left(t+2m_b^2-2m_t^2\right)+N\left(t_2+4m_b^2\right)+sq^2\right].
\end{eqnarray}
Here it is understood that all variables appearing on the right-hand sides are
to be taken as independent.
E.g., $N$ should be treated as independent of $t$, $u$, $w$, and $h$.
Notice that the UV divergences of $F_{+-}^1$ and $F_{+-}^2$ cancel in the
expression for ${\cal M}_{+-\lambda_W}$ in Eq.~(\ref{eq:m}).
Finally, we remark that we recover the SM result for $d\sigma(gg\to ZH)/dt$
due to one quark flavour \cite{kni} from Eq.~(\ref{eq:x}) by substituting
$m_W=m_Z$, $m_b=m_t$, and $\tan\beta=1$ and adjusting the strengths of the
axial-vector and Yukawa couplings.
In particular, the contribution proportional to the weak vector coupling then
vanishes as required by charge-conjugation invariance.
This serves as a useful check for our analytical and numerical analyses.

\vspace{1cm}
\noindent
{\bf Acknowledgements}
\smallskip

\noindent
B.A.K. thanks the KEK Theory Division for the hospitality extended to him during
a visit when this paper was prepared.
The work of A.A.B.B. was supported by the Friedrich-Ebert-Stiftung through
Grant No.~219747.
The II. Institut f\"ur Theoretische Physik is supported by the
Bundesministerium f\"ur Bildung und Forschung under Contract No.\ 05~HT9GUA~3,
and by the European Commission through the Research Training Network
{\it Quantum Chromodynamics and the Deep Structure of Elementary Particles}
under Contract No.\ ERBFMRXCT980194.

\end{document}